\documentclass{aa}

\usepackage{graphicx}

\begin{document}

\title{NTT infrared imaging of star cluster candidates towards the
central parts of the Galaxy.
\thanks{ESO proposals 67.B-0435 and 69.D-0225.}}

\author{C.M. Dutra\inst{1,2}, S. Ortolani\inst{3}, E. Bica\inst{4},
B. Barbuy\inst{1}, M. Zoccali\inst{5} and Y. Momany\inst{3} }

\offprints{C.M. Dutra -- dutra@astro.iag.usp.br}

\institute{Universidade de S\~ao Paulo, Instituto de Astronomia,
Geof\'\i sica e Ci\^encias atmosf\'ericas, CP\, 3386, S\~ao Paulo
01060-970, SP, Brazil\\ \mail{}
\and
Universidade Estadual do Rio Grande do Sul, Rua Bompland 512, S\~ao Borja
 97670-000, RS, Brazil\\
\and
Universit\`a di Padova, Dept.  di Astronomia, Vicolo dell'Osservatorio
2, 35122 Padova, Italy\\ \mail{}
\and
Universidade Federal do Rio Grande do Sul, Instituto de F\'\i sica,
CP\,15051, Porto Alegre 91501-970, RS, Brazil\\ \mail{}
\and
European Southern Observatory, Karl-Schwarzschild-Strasse 2, D-85748
Garching bei M\"unchen, Germany\\ \mail{}
}

\date{Received ; accepted }

\abstract{
 We address the issue  whether the central parts  of the Galaxy harbour
young clusters  other  than Arches, Quintuplet  and the  Nuclear Young
Cluster.  A large sample of centrally projected cluster candidates has
been recently identified from the 2MASS $J, H$ and  $K_s$ Atlas. We provide  a
catalogue of higher angular resolution and deeper images for 57 2MASS cluster candidates,
  obtained with the near-IR camera SOFI at the
ESO NTT telescope.  We classify 10 objects as  star clusters, some  of
them deeply embedded in gas and/or  dust clouds.  Three other objects
are probably star clusters, although the presence of dust in the field
does not exclude the possibility of their being field stars seen through low-absorption
regions.  Eleven  objects are concentrations of stars  in areas of little
or   no   gas,   and   are    classified   as    dissolving    cluster
candidates. Finally,  31 objects turned out to  be the blend  of a few
bright stars, not resolved as such in the low resolution 2MASS images.
By combining the above results with  other known objects we provide an
updated sample of 42 embedded clusters and candidates projected within
7$^{\circ}$.   As a
first  step we study Object  11 of Dutra \&   Bica (2000) projected at
$\approx 1^{\circ}$   from  the nucleus.     We  present $H$   and $K_s$
photometry and   study the  colour-magnitude   diagram and  luminosity
function.  Object 11 appears to be a less  massive cluster than Arches
or Quintuplet, and it is located at a distance from  the Sun $d_{\odot} \approx$
8 kpc,  with  a visual  absorption  $A_V \approx$   15. 
}

\titlerunning{NTT imaging of IR cluster candidates}

\authorrunning{Dutra et al.}

\maketitle

%

%________________________________________________________________

\section{Introduction}

The            2MASS         Atlas           (Skrutskie             et
al.     1997--http://pegasus.phast.umass.edu/2mass.html)  has    made
it possible    to study the infrared   population   of star clusters and
candidates towards the central part of the Galaxy (e.g. Dutra \& Bica
2000, 2001).  By central part  of the  Galaxy we  mean within
7$^{\circ}$ of the  nucleus (1 kpc at  the Galactic center  distance).
Recently, Portegies Zwart et al.  (2001) modeled cluster formation and
tidal survival  in a  more central  region within  1.43$^{\circ}$ (200
pc).

The central part of the Galaxy is known to  harbour the massive star
clusters Arches, Quintuplet and  the Nuclear Young cluster (e.g. Figer
et al.  1999a, Gerhard 2001 and  references therein). Several more may
exist according to simulations by Portegies Zwart  et al. (2001).
They estimated 50 massive clusters  within 200 pc, which would survive
to a  tidal  dissolution  time of   $\approx$ 70  Myr. A  fundamental
question is  whether clusters  predicted  by  Portegies Zwart  et  al.
(2001)   can be  detected.   Can young   clusters  such as Arches  and
Quintuplet be  detected not only within  200 pc but also  up to 1 kpc?
Are most central  clusters too  much absorbed  to be seen  at 2$\mu$m?
The samples of  Dutra \& Bica (2000,  2001) provided candidates within
both zones, which at the 2MASS angular resolution resembled the images
of Arches and Quintuplet as seen on the same material.  In the present
study  we employ larger  resolution    images to better select   these
samples, which is important for detailed  photometric studies with large telescopes.

Since   star clusters towards   the  central parts  dissolve in faster
timescales, we would expect to observe objects at different dynamical stages and differently
populated. Recently formed  clusters
will be embedded in the  parent molecular clouds in different  stages,
and  older  objects unrelated to gas  and  dust may  show  evidence of
advanced dynamical evolution.

\begin{figure}
\resizebox{\hsize}{!}{\includegraphics{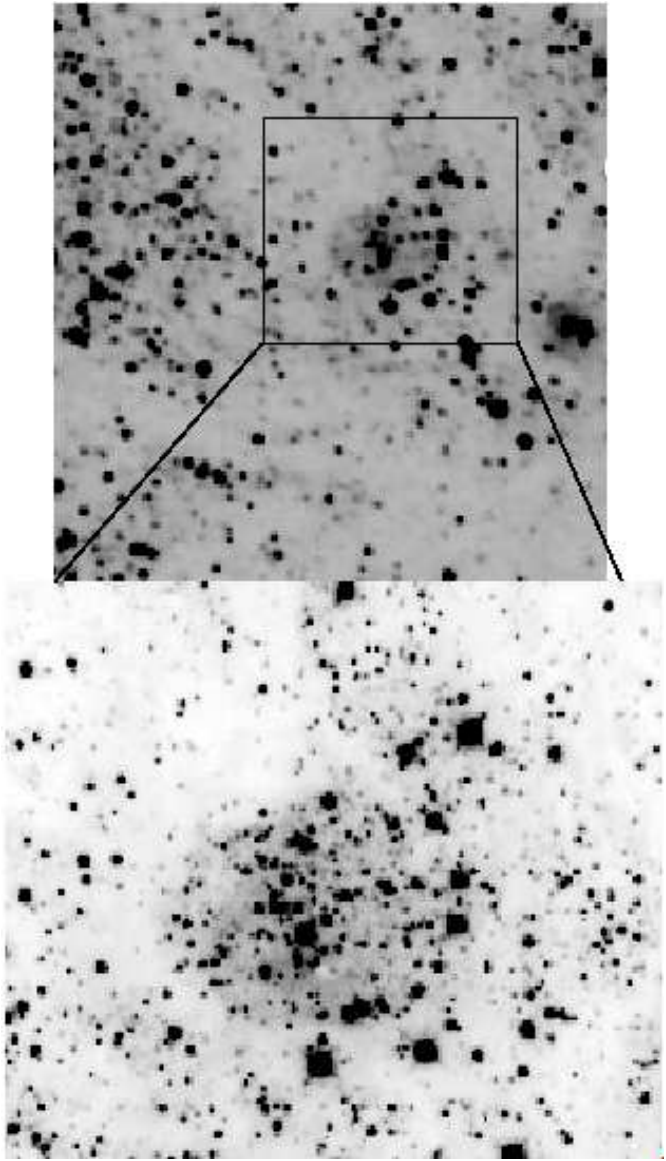}}
\caption{Images of object 11: a 5$^{\prime}\times 5^{\prime}$ K$_S$ 2MASS extraction
in the upper panel and a 2$^{\prime}\times 2^{\prime}$ K$_S$ SOFI large field extraction in the lower panel. North is up and west to the right.}
\label{fig3}
\end{figure}

Using the $J,H$ and $K_S$ 2MASS images, Dutra \& Bica (2000, hereafter
Paper I) detected 58 infrared  star cluster candidates projected  near
the Centre.   Typical dimensions  were  1-2$^{\prime}$, like  those of
Arches  and Quintuplet.   At higher  resolution the  Arches cluster is
concentrated  while the  Quintuplet   cluster is loose  (Figer  et al.
1999a). Seven  additional   candidates  in  the central    parts  were
indicated by Dutra  \& Bica (2001,  hereafter Paper II). The  angular
resolution of   the 2MASS Atlas is  not  high, and it  is necessary to
increase it and obtain deep images in  order to constrain the nature of
the candidates. This is  the   objective of  the present  $K_{\rm  s}$
survey with the  3.55 m ESO New Technology  Telescope (NTT) making use
of the resolving power of the SOFI camera  to study 52 candidates from
Paper I and 5 from Paper II.  We also present detailed photometry of
Object 11 from Dutra \& Bica (2000).

In Sect. 2 we present the $K_{\rm  s}$ survey and  results. In
Sect. 3   we present the  $H$ and  $K$ photometry  of Object  11 and
analyse   the results.  Finally,  concluding remarks   are given  in
Sect. 4.

\section{A $K_{\rm s}$ survey of cluster candidates}

In view of selecting the previous  2MASS samples for detailed studies
with large telescopes we present a  $K_{\rm s}$ imaging survey carried
out with the NTT for 57 cluster candidates.

\subsection{Observations and Reductions}

We employed the SOFI camera at the NTT Nasmyth A focus with the detector  Rockwell 
Hg:Cd:Te of 1024x1024 pixels (18.5 $\mu$m) 
Hawaii array.  We used the Small Field mode 
(2.47$^{\prime}\times$2.47$^{\prime}$ and scale 0.145$^{\prime\prime}$/pixel) 
on June 27, 2001 and the  Large Field mode 
(4.94$^{\prime}\times$4.94$^{\prime}$ 
and scale 0.292$^{\prime\prime}$/pixel)
 on July 4 - 6, 2002. 
 The K$_S$ band (2.162$\mu$m) allows one to minimize dust absorption effects 
(A$_{K_{\rm s}}$=0.11A$_V$, Cardelli et al. 1989).
 Owing to weather conditions only a few objects were observed on 
 June 27, 2001 and July 5, 2002. 
In June 2001 we adopted  a detector integration time DIT = 4 sec, a
 number of detector integrations  NDIT = 5 and a number of exposures 
NEXP = 20, whereas on July 2002  DIT = 7,
 NDIT = 7 and NEXP = 18.  
More details on the observations are given in Table 1. 

For infrared observations it is necessary to frequently subtract the sky 
thermal emission. We used the subtraction technique for small objects or uncrowded fields, from the SOFI Users 
Manual (Lidman  et al.  2000). The reduction      consisted  of dark frame
subtraction, sky subtraction  and flat   fielding, following the
 steps given in the SOFI manual.

 In the process of flat fielding  the illumination correction  
frames and the bad pixels maps,
both available from the ESO webpages, were used.

\begin{table*}
\caption{\scriptsize
Log of observations.}
\begin{tabular}{lccccc}
\hline\hline
Date&Objects&Filter&Exposure&Seeing&Field\\
\hline
&&&(min.)&($^{\prime\prime}$)&\\
\hline
June 27 2001&3, 19, 24, 26, 01-40&H,K$_s$&6.7&1.5&Small\\
July 4 2002&1, 2, 4, 5, 6, 7, 8, 9, 10, 11, 12, 13, 14, 15, 16, 17, 20, 21,&K$_s$&14.7&0.8&Large\\
July 4 2002&22, 23, 25, 26, 29, 40, 41, 42, 43, 44&K$_s$&14.7&0.8&Small\\
July 5 2002&30, 31, 32&K$_s$&14.7&1.5&Large\\
July 6 2002&01-01,01-02,01-41,01-42, 33, 34, 35, 36, 37, 38, 39,&K$_s$&14.7&0.8&Large\\
July 6 2002&46, 47, 48, 49, 52, 53, 54, 55, 56, 57, 58&K$_s$&14.7&0.8&Large\\
%July 6 2002&11&H,K$_s$&15.0&0.8&Small\\
\hline
\end{tabular}
\begin{list}{}
\item  Notes: Objects from Dutra and Bica (2001) are indicated by 01-, else  from Dutra \& Bica
(2000).
\end{list}
\end{table*}

\subsection{Results}

The higher angular resolution and depth of the NTT images with respect
to the 2MASS Atlas allowed us  to classify the objects more clearly now. 
Note that the classifications are based on eye estimates of the stellar overdensity on the images.

\subsubsection{Confirmed Clusters}

Objects 11 (Fig. 1), 52 (Fig. 2a), 6  (Fig. 2b), 5 (Fig. 2b), 55, 10 (Fig. 2c), 01-40 and 01-41 (Fig. 2d) appear to be  resolved
star clusters, most of them embedded in nebulosity.

 Objects 10 and 11 have counterparts in the Mid Space Experiment (MSX) survey (Egan et al. 1999). This infrared survey 
provides data and images in the bands $A$ (8.28$\mu$m), $C$ (12.13$\mu$m),  $D$ (14.65$\mu$m)
and $E$ (21.3$\mu$m) and is electronically available at the Web site {\rm http://irsa.ipac.caltech.edu/applications/MSX/}. 
 The infrared emission of Objects 11 and 10 in these MSX bands is probably due to
dust heated by massive stars. In addition Object 10 has a counterpart in the IRAS point source catalog, [IRAS\,17470-2853].
Using the colours diagram of IRAS PSC sources associated with ultra compact HII region from Wood \& Churchwell (1989) and 
 IRAS\,17470-2853's 12$\mu$m, 25$\mu$m and 60$\mu$m IRAS fluxes, we find that it is an ultracompact HII region.  In the field of Objects 11/10 there is diffuse emission  with diameter $\approx$5$^{\prime}$ in a Digitized Sky Survey $R$ band image ({\rm http://cadcwww.dao.nrc.ca/cadcbin/getdss}), corresponding to the 
optical HII region Sh2-21 (Sharpless 1959). Using the 2MASS Atlas one can trace in the area  a dust (molecular) cloud with a  diameter of $\approx$15$^{\prime}$. These structures, if located at about the Galactic center distance, would define a giant
molecular cloud and HII region with 2 embedded star clusters.  In Sect. 3 we provide a NTT $H$ and $K$ photometry analysis of Object 11.

\subsubsection{Possible Clusters}

Objects 26 (Fig. 3a) and 56 (Fig. 3b) appear to be star clusters deeply embedded in dust and gas, in very 
early  stages  of star formation.

Near objects 12, 58 (Fig. 4a) and 01-42 (Fig. 4b) we find 
dust  absorption. Therefore, it  is  not excluded that
these objects  are  field stars seen through low absorption windows.
Infrared colour-magnitude  diagrams (CMD) may clarify the issue.

Object  01-01 (Fig. 4c) is  an open  cluster candidate. It is an
interesting target for CMD studies, since not much is known about open
clusters towards the Galactic center, a  few kpc away from  the Sun. 

Object 49 is a clump of stars embedded in a nebula.

Several  objects turned out to  be concentrations of stars with little
or no gas/dust in the area. They may be clusters in the
process  of dissolution. In  the solar neighbourhood the timescale for
dissolution of open  clusters is a  few Gyr or  less and several have
been studied  in detail (Pavani et  al. 2001,  Carraro 2002, Pavani et al. 2003), while in
the central  200 pc  the timescale is  reduced  to $\approx$ 70  Myr (Portegies
Zwart et al. 2001). These candidate dissolving clusters are objects 1, 7, 17, 25, 31, 32,
35, 40, 41, 42 and 01-02. Object 1 is shown in Fig. 4d.

\subsubsection{Spurious Identifications}

Based on   NTT images,  the following   objects were  found not to  be
clusters, but one or more relatively bright stars (plus faint ones) or
clumps which were previously unresolved   in the 2MASS images   (Paper
I). These are the objects 2, 3, 4, 8, 9,  13, 14, 15,  16, 19, 20, 21,
22, 23, 24, 29, 30, 33, 34, 36, 37, 38, 39, 43, 44, 46, 47, 48, 53, 54
and 57. Recording   such blended images which  mimic  clusters is also
important in  view of future systematic cluster  surveys  on the 2MASS
Atlas.

\begin{figure}
\resizebox{\hsize}{!}{\includegraphics{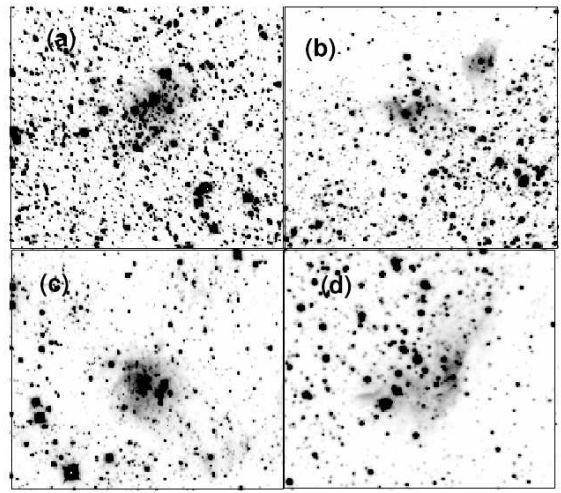}}
\caption[]{$K_s$ SOFI large field extractions: (a) 2$^{\prime}\times 2^{\prime}$  of Object 52, 
(b) 2$^{\prime}\times 2^{\prime}$  of Object 6 (centre) and Object 5 (right), (c) 2$^{\prime}\times 2^{\prime}$  of Object 10,
and (d) 2$^{\prime}\times 2^{\prime}$  of Object 01-41. North is up and west to the right.}
\label{fig1}
\end{figure}

\begin{figure}
\resizebox{\hsize}{!}{\includegraphics{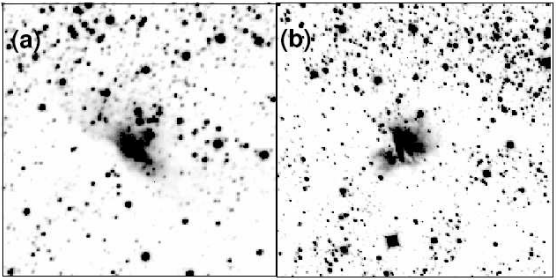}}
\caption[]{$K_s$ SOFI large field extractions: (a) 2$^{\prime}\times 2^{\prime}$  of Object 26 and 
(b) 2$^{\prime}\times 2^{\prime}$  of Object 56. North is up and west to the right.}
\label{fig1}
\end{figure}

\begin{figure}
\resizebox{\hsize}{!}{\includegraphics{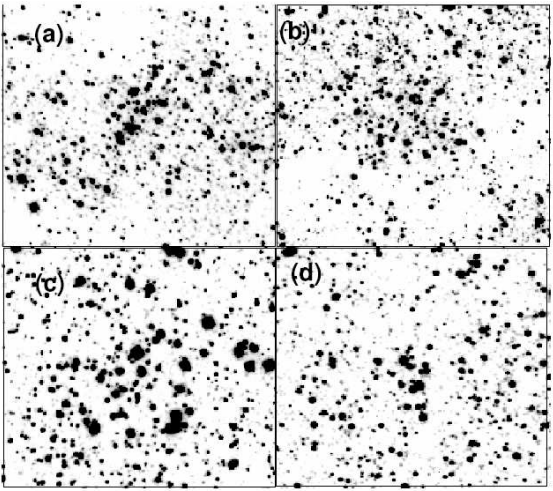}}
\caption[]{$K_s$ SOFI large field extractions: (a) 2$^{\prime}\times 2^{\prime}$  of Object 58, 
(b) 3$^{\prime}\times 3^{\prime}$  of Object 01-42, (c) 3$^{\prime}\times 3^{\prime}$  of Object 01-01,
and (d) 2$^{\prime}\times 2^{\prime}$  of Object 1. North is up and west to the right.}
\label{fig1}
\end{figure}

\subsubsection{Relation to nebulae}

We checked the possibility of  association of the present objects with
optical and radio   nebulae (e.g.  Kuchar  \& Clark  1997;  Caswell \&
Haynes 1987; Lockman 1989), which  in turn reinforces the  possibility
that they are young  stellar  systems.  Object 58   appears to be related to  the
optical HII region Sh2-17, Objects 5 and 6 to Sh2-20,
and Objects  7, 10, 11  and 12 to Sh2-21.   Although Object 01-01 is
projected close to the dark  nebula LDN74 (Lynds  1962), it appears to
be an evolved open cluster and consequently unrelated to it.

Object 26  is related to  the nuclear star-forming  complex  Sgr D and
Object 52  to Sgr E  (Liszt 1992).  Object 01-40 is   in the radio HII
region  G353.4-04, 01-41 in  G354.664+0.470  and 01-42  in G359.3-0.3.
Object 55 appears to be related to G359.54+0.18, and 56 to G359.7-0.4.

\subsection {The updated sample of central clusters and candidates}

We show in Fig. 5 the angular distribution  including the results from
the present survey.  We   indicate  clusters, cluster candidates   and
cluster  dissolution candidates.    The  overall sample spans  objects
within  7$^{\circ}$ (1 kpc  at the  Galactic  Center distance), and we
also indicate   the region of  200  pc modeled by  Portegies  Zwart et
al. (2001). In addition to objects from Dutra \&  Bica (2000, 2001) we
show objects   from a recent  2MASS  cluster search in the  directions of
optical and radio nebulae (Bica  et al.  2003).   The asymmetry in the
sense of more  objects projected  on the eastern  part of  the central
Galaxy is due  to the fact that the  eastern side has been  surveyed
with   2MASS for infrared  clusters in  all  directions (Dutra \& Bica
2000), while  the western side has  been  mostly surveyed for embedded
clusters in the directions of nebulae.

Since we are dealing mostly   with embedded clusters and   candidates,
these objects  constitute an important sample  for probing the cluster
populations related to intervening  spiral arms and the central  parts
of the Galaxy.  We conclude that  42 objects are now available within
7$^{\circ}$ of  the center, being  19 of  them projected on  Portegies
Zwart et al.'s zone.

\begin{figure*}
\resizebox{\hsize}{!}{\includegraphics{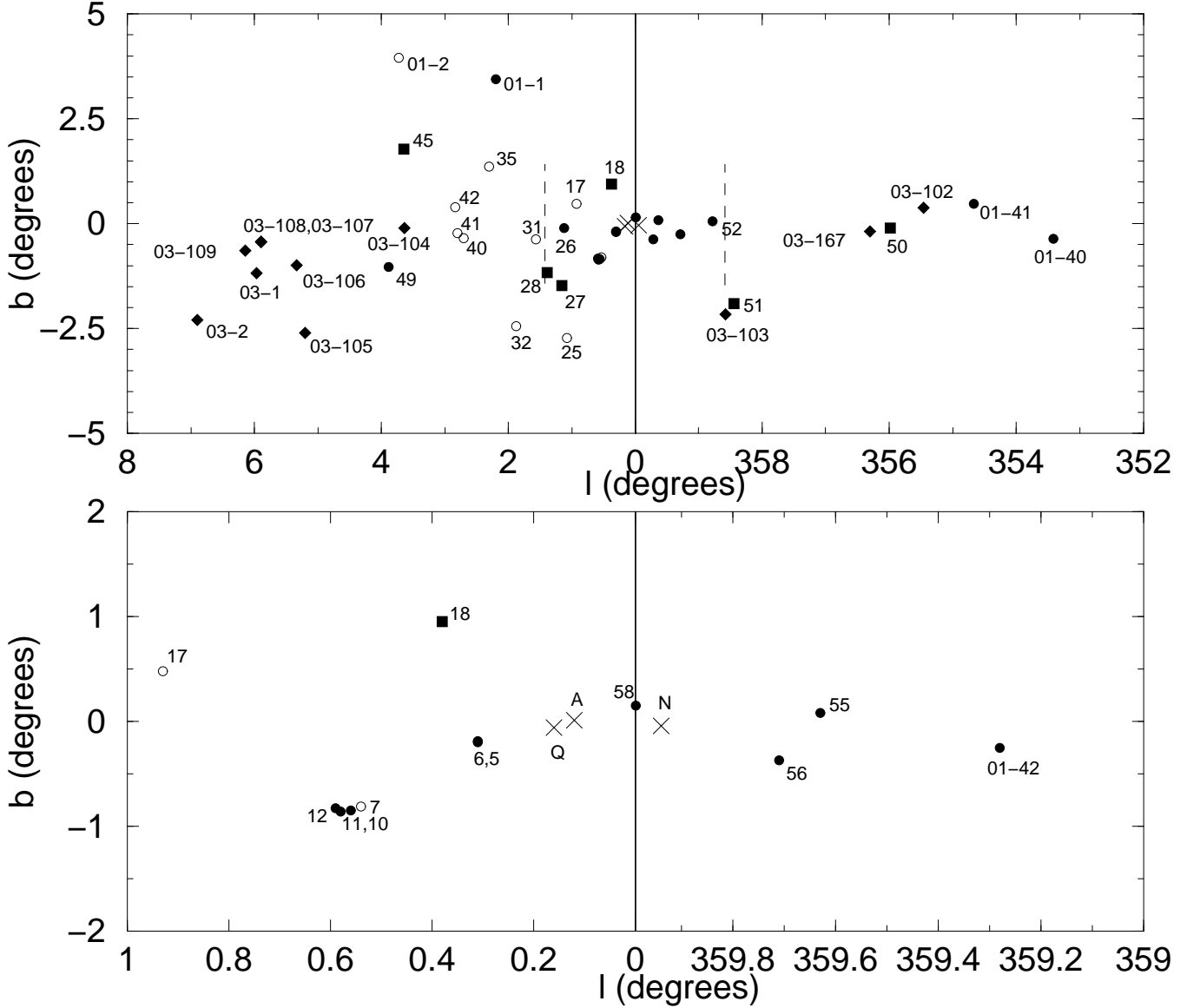}}
\caption{Upper panel: updated angular distribution of star clusters and
candidates in   the inner 7$^{\circ}$  (1 kpc at  the Galactic Center
distance). Lower  panel:   blowup  to  show  in  detail    the central
region within 1$^{\circ}$.  Objects from Dutra \& Bica (2001) are labeled
as  01- and from  Bica et al.   (2003)  as 03-, the other objects are from
Dutra \& Bica (2000). Crosses are the Arches, Quintuplet and Nuclear Young
Cluster.  Filled circles are objects  from the present study which are
confirmed   as clusters or   candidates.  Open circles  are dissolving
cluster candidates.  Filled squares are   objects from Dutra  \&  Bica
(2000)  not included in the  present work. Filled diamonds are objects
from Bica et al. (2003).}
\label{sample}
\end{figure*}

\section{NTT $H$ and $K$ photometry of Object 11}

The  only  object for which two    colour photometry was  available is
Object 11  from  Dutra \& Bica  (2000).    It is located   at $\approx
1^{\circ}$ ($\ell =$  0.58,  $b =$  -0.86) from  the Galactic nucleus,
thus  projected    on the   zone     modeled by  Portegies   Zwart  et
al. (2001). In this section we present a more detailed analysis of the
cluster population, using its $H,K_s$ CMD.

\subsection{Infrared data}

The $H$ and $K_{\rm  s}$ observations at  the NTT were
obtained with  the SOFI camera equipped with   the Hawaii $1024 \times
1024$ HgCdTe detector, with a pixel size of 18.5$\mu$m, on July 2002.

The   observations     used the    SOFI   small    mode,  with   scale
$0.145^{\prime\prime}$/pixel       and  field     $2.47^{\prime}\times
2.47^{\prime}$.   The detector integration times  DIT were $30$ and $20$
seconds  for $H$ and  $K_{\rm s}$, respectively. Similarly, the number
of detector integrations NDIT  was $2$ and $3$,  for a total number of
exposures NEXP of $15$.

The thermal emission subtraction technique   follows Lidman et
al.  2000).  The standard stars  9150, 9157  and  9170 from Persson et
al. (1998) were  observed  at different airmasses.  For  each standard
star five measurements of 2 sec for each airmass were obtained.

Dark frame  subtraction, sky   subtraction   and flat fielding    were
applied.   We used illumination correction  frames  and the bad pixels
masks, available from the ESO webpages for flat fielding.

We used a $K_{\rm s}$ filter, and we consider  that this introduces an
extra uncertainty   in  the $K$  filter calibration  of  $\pm 0.02$ mag
(e.g. Ivanov et  al.  2000).    The  instrumental magnitudes  of   the
standard stars  were normalized to $1$  sec exposure and zero airmass,
according to the following equation:

\begin{equation}
m^{'}=m_{\rm ap}+2.5\, \log(t_{\rm exp})-K_{\lambda }\, X \,
\end{equation}

\noindent where $m_{\rm ap}$   is the  mean  instrumental magnitude   of the $5$
measurements in  a circular aperture  of radius $R=5.2$ arcsec, $X$ is
the  mean  airmass  and $t_{\rm exp}$  is   DIT in  seconds.  The mean
extinction coefficients adopted   for La Silla  are:  $K_{H}=0.03$ and
$K_{K}=0.05$  (from  ESO  webpages).    A least squares fit of  the
normalized instrumental magnitudes to the magnitudes of Persson et al.
(1998) gave the following relations:

%%%%%%%%%%%%%%%%%%%%%%%%%%%%%%%%%%%%%%%%%%%%%

\begin{equation}
H-h_{\rm s}=-0.018\,\times (H-K)\,+\, 22.920\,
\end{equation}

\begin{equation}
K-k_s=-0.005\,\times (H-K)\,+\,22.35\,
\end{equation}
%%%%%%%%%%%%%%%%%%%%%%%%%%%%%%%%%%%%%%%%%%%%

The r.m.s.   scatter of   the residuals  of  the  fit is  $0.025$ and
$0.021$ mag in $H$  and $K_s$  respectively.  Together  with the
$K_s$ to $K$ filter,  this yields total uncertainties of $0.025$
and $0.041$ in the $H$ and $K_s$ bands respectively.

The photometric extractions were performed on the stacked images using
DAOPHOT/ALLSTAR package.   The final calibrated  magnitudes took  into
account aperture corrections applied   to PSF magnitudes ($0.115$  and
$0.080$ mag for $H$ and $K_{\rm s}$ respectively).

\begin{figure}
\resizebox{\hsize}{!}{\includegraphics{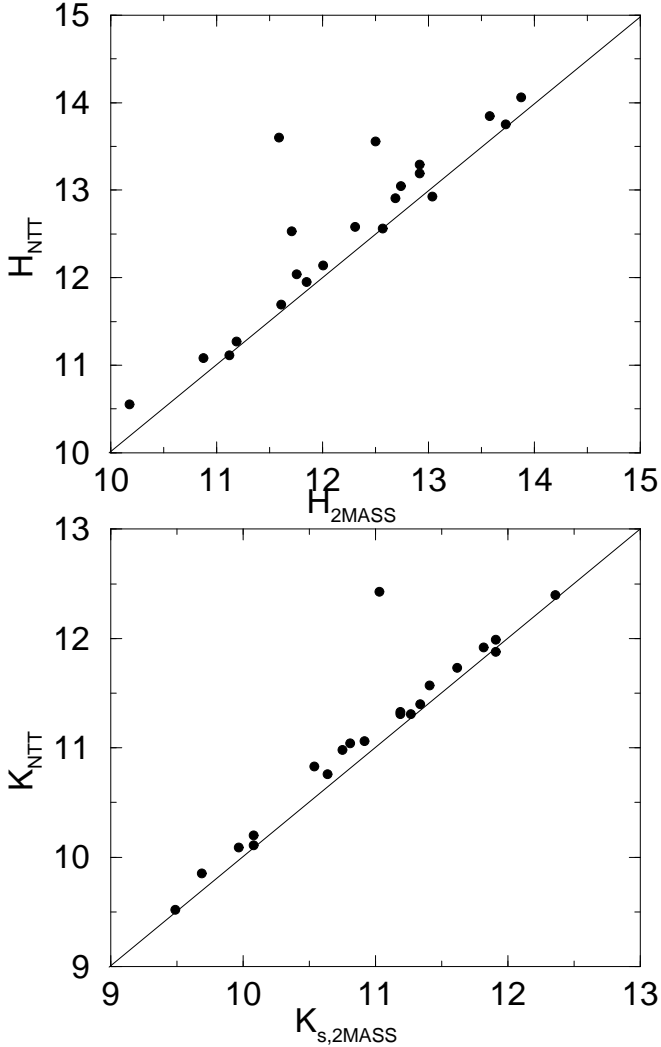}}
\caption{$H$ and $K$ comparisons of the present and 2MASS photometries.
Identity functions  are shown.}
\label{sample}
\end{figure}

\subsection{Comparison with 2MASS}

We searched in our observations for stars in common with the 2MASS
photometry. They occur both in the object area and in the surrounding
field, reaching $H\approx14$ and $K_s\approx12.5$. The comparison is
shown in Fig. 6. There is a linear relation between the photometries,
which departs somewhat from identity. The zero-point differences
between linear fits to the data and the identity function are 0.17 
for $H$ and 0.11 for $K-K_s$. In the linear fits we disregarded three
deviant points in $H$ and one in $K$ which are probably due to, e.g.,
crowding, cosmic rays or variable stars.

\subsection {Analysis}

\begin{figure}
\resizebox{\hsize}{!}{\includegraphics{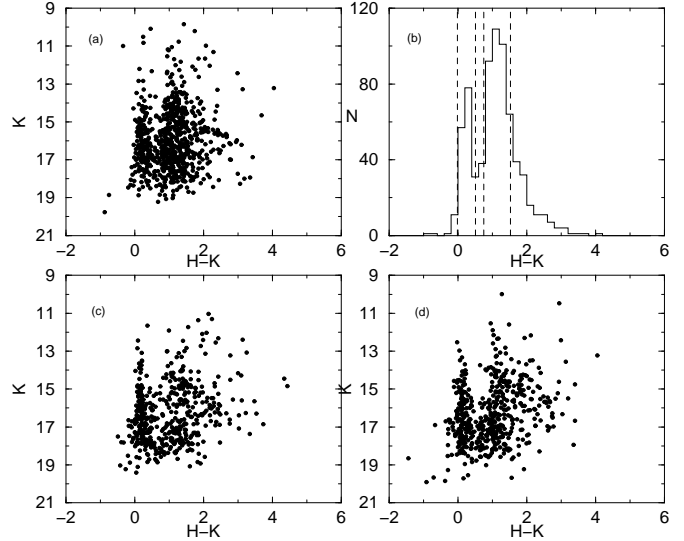}}
\caption{$K, H-K$ diagrams for cluster (panel a) and fields 1 (panel c)
and 2 (panel d) extractions. Panel b shows $H-K$ colour histogram for
the object area.}
\label{sample}
\end{figure}

\begin{figure}
\resizebox{\hsize}{!}{\includegraphics{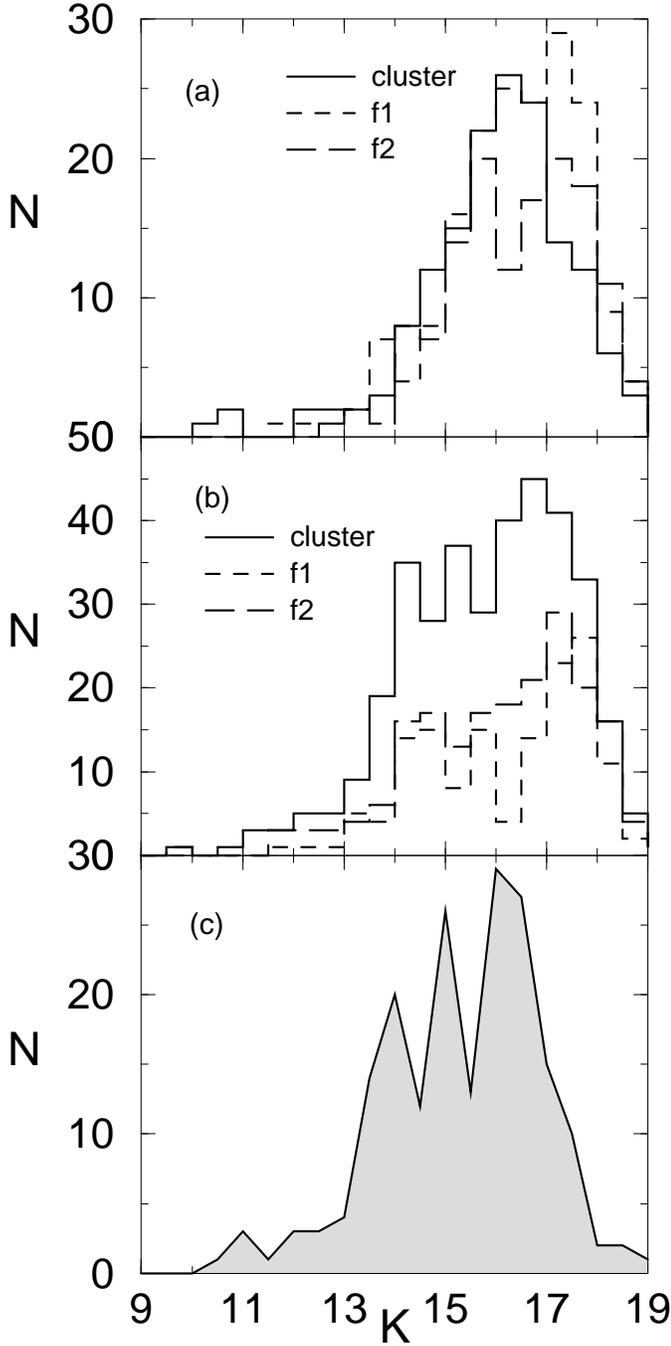}}
\caption{Upper panel: $K$ histogram for stars in the range 0.0 $<H-K<$ 0.5.
Middle panel: $K$ histogram for stars in the range 0.75 $<H-K<$ 1.5.
Continuous line corresponds to object while dashed and long dashed lines
to the fields. Lower panel: $K$ cluster luminosity function  derived from
average field subtraction for 0.75 $<H-K<$ 1.5.}
\label{sample}
\end{figure}

For the photometric analyses we selected three 1.2$^{\prime} \times$1.2$^{\prime}$ square
regions in the object area and  two fields on  either side to the east
(Field 1) and west (Field 2).

Fig. 7 shows $K,  H-K$ CMDs from  the NTT  photometry for the  object
(panel a) and the  two side fields  (panels c and d).  In the CMDs two
sequences are clearly present, which peak at $H-K
\approx$ 0.25 and 1.10 respectively,  as shown in the colour histogram
of  panel b). Note that in  the object CMD  the second sequence is more
populated than  in the CMDs of the  fields.   In order to  verify this
point which would  indicate   the presence  of  a  cluster,  we  built
histograms of the   distribution  of stars along   the  $K$ magnitude,
assuming a $H-K$ colour interval for each sequence.   Fig. 8 shows the
results,    where  we assumed  for the    blue  and  red sequences the
respective  colour  intervals  $0.0<H-K<0.5$ and  $0.75<H-K<1.5$.  The
upper panel of Fig. 8 shows that no  contrast is observed for the blue
sequence, indicating that we are dealing with field  stars both in the
object  and field extractions.  The  middle panel  of  Fig. 8 shows a
strong contrast between the object  and field extractions for the red
sequence, especially  for stars fainter  than $K  =$  13, indicating a
cluster.  Finally, in the   lower panel of   Fig. 8 we show  the field
subtracted luminosity function which increases to $K \approx$ 16, also
suggesting  a cluster. Completeness effects  must be affecting fainter
magnitudes.

For a more detailed CMD analysis we statistically 
decontaminated the cluster's CMD (Fig. 9a) from the Bulge stellar field contribution represented by
stars extracted outside the cluster area (Fig. 9b). In the decontamination procedure, we scaled the Bulge stars contribution considering the cluster area.
 For each star in the scaled bulge CMD we picked the closest star in the cluster's CMD, and subtracted it. The distance on the CMD from bulge field star and star in the cluster area was defined as:
 $$d=\sqrt{[7 \times \Delta(H-K)]^2 + \Delta H^2}$$(see Zoccali et al. 2003, for more details about decontamination procedure).  Figure 9c 
  shows the decontaminated cluster CMD, while the CMD of the stars statistically removed from the cluster area CMD is shown in Fig. 9d.

The bulge tilted horizontal branch can be seen in Fig. 9b starting 
from $K =$ 13.8 and $H-K =$ 0.76. Assuming that the dust/molecular cloud must
be behind the bulk of the bulge population seen in the diagram (it must have a much higher extinction), we can set a lower limit for
the distance of Object 11. We use as reference for the bulge field population
that of the CMDs at the minor axis b = -6{$^{\circ}$} (Zoccali et al. 2003).
In this field the bulge clump is located at $K =$ 12.7 and $H-K =$ 
0.25. The reddening at -6{$^{\circ}$} is $E(B-V) =$ 0.37, corresponding to
$E(H-K) =$ 0.10 and A$_K =$ 0.175. The intrinsic value is $K_0 =$ 12.53,
and $(H-K)_0 =$ 0.15. The colour difference between the bulge in the
direction of Object 11 and the -6{$^{\circ}$} field is $\Delta (H-K) =$ 
0.60 corresponding to $A_K =$ 1.05 (or $A_V =$ 9.6). The relative distance modulus is
$\Delta(m-M)_0 =$ 0.2.
Given the distance uncertainties this means that the two fields are at the
same distance, i.e. compatible with that of the Galactic center.
From the colour excess of the Object 11 field we get the minimum absorption 
which is likely the absorption in front of the cloud ($A_V =$ 9.6). This value is consistent
with the presence of an optical HII region (Sect. 2.2).

The 10$^{th}$ brightest star method to  determine cluster distances is
useful  when no  spectral  type  information  is available  for  young
clusters  (Dutra  \&  Bica 2001). The    method assumes similarity  of
luminosity functions of two clusters, and  avoids uncertainties due to
the brightest  stars, for which  luminosity effects are important.  The
10$^{th}$ brightest star in the giant H\,II region cluster NGC\,3603 is
an O4V star (Moffat 1983).  We will make  two assumptions, that Object
11 is:  (i) a massive cluster where  the 10$^{th}$ brightest star is an
O5V, or (ii) a less massive  cluster where the 10$^{th}$ brightest star
would correspond to a B0V star. According to  Cotera et al. (2000) O5V
stars  have   intrinsic   magnitude    and    colour $M_K=-4.81$   and
$(H-K)_0=-0.08$, while B0V stars have $M_K=-3.34$ and $(H-K)_0=-0.07$.

Object  11 (Fig. 1)   is embedded in  nebulosity  which suggests very
young ages  ($t<5$ Myr).  For the subsequent  analysis we use Cotera  et al.'s (2000)  values for
upper Main  Sequence (MS) stars and the 10$^{th}$ brightest star method to estimate the reddening and
distance parameters for Object 11. We  derive from Object  11's decontaminated CMD (Fig. 9c) a reddening
$E(H-K)=$ 1.14 with an important differential reddening $\delta E(H-K)
= \pm 0.35$ as  can be estimated from a mean reddening of $H-K =$ 1.07 in comparison to Cotera  et al.'s (2000) MS stars. Adopting the
absorption ratios  from Schlegel et   al. (1998), one can  derive  the
relation $A_K=1.436   E(H-K)$, which leads  to  $A_K=1.64$. This value
corresponds to a visual absorption $A_V \approx$ 15.

According to Fig. 9c  the 10$^{th}$ brightest star  in  Object 11  is
located at $K=12.7$.  Assuming Object 11 as  a massive cluster similar
to NGC\,3603,  we  derive a distance  from the  Sun  $d_{\odot}= 14.9$
kpc. On the other hand, we assume a less massive cluster with a B0V star as 10$^{th}$
brightest, then $d_{\odot}=7.6$ kpc. The Quintuplet cluster,
argued to be at  the Galactic centre distance,  has $A_V=$ 29 (Figer et
al. 1999b), much  higher than that of  Object 11. Therefore, Object 11
appears to be slightly in the foreground  of the Galactic  centre, being a less
reddened   and  less   massive   cluster,  favouring    the   solution
$d_{\odot}=7.6$ kpc.  Differential reddening  is  the main  source  of
uncertainties in   distance,  assuming that   the   spectral type   of
 the 10$^{th}$ brightest star is correct. For Object 11 $\delta E(H-K) = \pm
0.35$ implies  $\delta A_K = \pm  0.50$ and a distance  uncertainty of
$\pm$ 1.2 kpc. Figure 10 shows the decontaminated  Object 11 CMD 
superimposed on Cotera  et al.'s (2000) MS
spectral type distribution considering
the B0V   reddening/distance solution.  Note    that we are   probably
reaching F0 stars.  It would be important to carry out spectroscopy of
cluster stars in order to further constrain the distance.

\begin{figure}
\resizebox{\hsize}{!}{\includegraphics{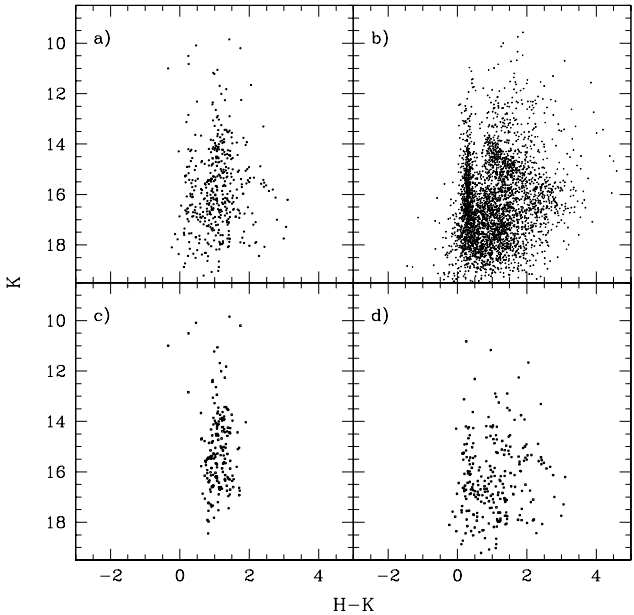}}
\caption{$K, H-K$ diagrams: (a) Object 11 area, extraction of stars with distance $r<$ 43.5$^{\prime\prime}$; (b) Bulge field, extraction of stars outside where field stars $r>$ 58$^{\prime\prime}$;
(c) clean cluster CMD; and (d) stars subtracted from the Object 11 area CMD in order to obtain the decontaminated CMD.}
\label{sample}
\end{figure}

\begin{figure}
\resizebox{\hsize}{!}{\includegraphics{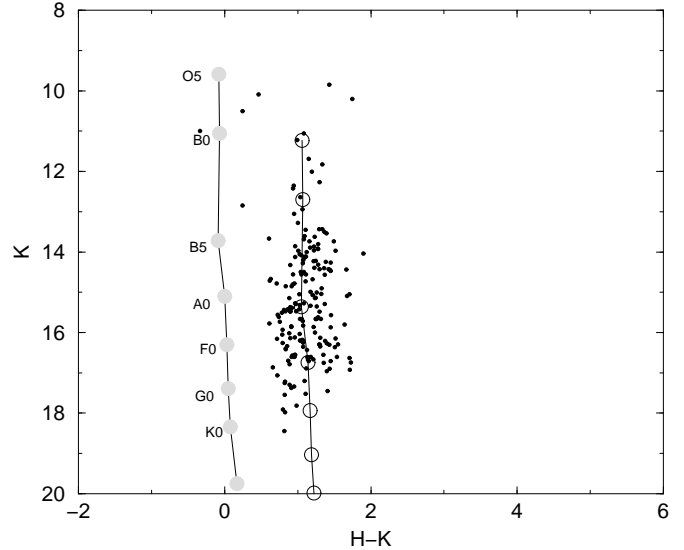}}
\caption{$K, H-K$ diagram of object, where field stars have been statistically
subtracted. MS spectral types sequence from Cotera et al. (2000): reddening-free and at 7.6 kpc (gray circles and solid line) and with a reddening/distance solution with $A_K=$ 1.64
($A_V=$ 15) and $d_{\odot}=$ 7.6 kpc (open circles and solid line).}
\label{sample}
\end{figure}

\section{Concluding remarks}

This study aimed to settle a star cluster sample for the exploration of the
structure and star formation events towards the central parts of the Galaxy.  
By  means of ESO NTT observations  we increased the angular resolution
and depth of observations  for 57 infrared cluster candidates  towards
the  central Galaxy, as  compared to those  of the  2MASS Atlas, where
these objects had  been first  identified.   We explored  this sample
showing which  objects are clusters,  remain as cluster  candidates or
appear  to be dissolving   cluster   candidates.  We also    indicated
additions from  other samples giving an updated  census of  42 objects
within 7$^{\circ}$ (1 kpc distance from the Galactic centre), of which
19  are within 1.43$^{\circ}$  (200  pc).  Detailed photometry of  the
present set  of objects using large telescopes  is  required to derive
their properties such as star membership, reddening  and age, in order
to  establish their location in the  Galaxy as intervening spiral arms
or the Galactic Center.

 We analyzed in detail one  of these objects projected at $\approx
1^{\circ}$ of the nucleus. $H$ and $K$  photometry of Object 11 showed
a colour-magnitude diagram and  luminosity function of a  cluster. The
cluster has $A_V=15$ and appears to be located at $d_{\odot} \approx 8
\pm 1.2$ kpc from the Sun, therefore not far from the Galactic centre. The
cluster appears to be less massive than the Arches and  Quintuplet clusters.

\begin{acknowledgements}

We   acknowledge  support from  the  Brazilian   Institutions CNPq and
FAPESP. CMD  acknowledges     FAPESP  for   a  post-doc     fellowship
(proc.    2000/11864-6).    SO   thanks    the  Italian      Ministero
dell'Universit\`a e della  Ricerca  Scientifica e Tecnologica  (MURST)
under the program  on   'Stellar Dynamics  and Stellar  Evolution   in
Globular Clusters: a Challenge for New Astronomical Instruments'.

\end{acknowledgements}


\begin{thebibliography}{}
\bibitem[]{} Bica, E., Dutra, C.M., Soares, J.B. \& Barbuy, B., 2003, A\&A, submitted.
\bibitem[]{} Cardelli, J.A., Clayton, G.C. \& Mathis, J.S. 1989, ApJ, 345, 245
\bibitem[]{} Carraro, G. 2002, A\&A, 385, 471
\bibitem[]{} Caswell, J.L. \& Haynes, R.F., 1987, A\&A, 171, 261
\bibitem[]{} Cotera, A.S., Simpson, J.P., Erickson, E.F., et al.  2000, ApJS, 129, 123
\bibitem[]{} Dutra, C.M. \& Bica, E. 2000, A\&A, 359, L9
\bibitem[]{} Dutra, C.M. \& Bica, E. 2001, A\&A, 376, 434
\bibitem[]{} Egan, M.P., Price, S.D., Moshir, M.M., et al. 1999, The Midcourse Space Experiment
Point source Catalog Version 1.2 Explanatory Guide, AFRL-VS-TR1999-1522, Air Force Research Laboratory
\bibitem[]{} Figer D.F., Kim S.S., Morris M., et al. 1999a, ApJ, 525, 750
\bibitem[]{} Figer D.F., McLean, I.S. \& Morris, M.  1999b, ApJ, 514, 202
\bibitem[]{} Gerhard, O. 2001, A\&A, 546, L39
\bibitem[]{} Girardi, L., Bertelli, G., Bressan, A., et al. 2002, A\&A, 391, 195
\bibitem[]{} Ivanov, V.D., Borissova, J. \& Vanzi, L. 2000, A\&A, 362, L1
\bibitem[]{} Kuchar, T.A. \& Clark, F. O. 1997, ApJ, 488, 224
\bibitem[2000]{lidman00} Lidman, C., Cuby, J-G.~\& Vanzi, L. 2000, in ``SOFI user's manual'' Doc. No. LSO-MAN-ESO-40100-0003, issue 13
\bibitem[]{} Liszt, H.S. 1992, ApJS, 82, 495
\bibitem[]{} Lockman, F.J., 1989, ApJS, 71 469
\bibitem[]{} Lynds, B.T. 1962, ApJS, 7, 1
\bibitem[]{} Moffat, A.F.J. 1983, A\&A, 124, 273
\bibitem[]{} Pavani, D.B., Bica, E., Dutra, C.M., et al.  2001, A\&A, 374, 554
\bibitem[]{} Pavani, D.B., Bica, E., Ahumada, A.V., et al.  2003, A\&A, 399, 113 
\bibitem[]{} Persson, S.E., Murphy, D.C., Krzeminski, W., et al.  1998, AJ, 116, 2475
\bibitem[]{} Portegies Zwart, S.F., Makino, J., McMillan S.L.W., et al.  2001, ApJ, 546, L101
\bibitem[]{}  Schlegel, D.J., Finkbeiner, D.P. \& Davis, M. 1998, ApJ, 500, 525
\bibitem[]{} Sharpless, S. 1959, ApJ, 4, 257
\bibitem[]{} Skrutskie, M., Schneider, S.E., Stiening, R., et al. 1997, in {\it The Impact of Large Scale Near-IR Sky Surveys}, ed. Garzon et al., Kluwer (Netherlands), 210, 187
\bibitem[]{} Stetson, P.B. 1987, PASP, 99, 191
\bibitem[]{} Stetson, P.B. 1994, PASP, 106, 250
\bibitem[]{} Wood, D.O.S. \& Churchwell E. 1989, ApJ, 340 265
\bibitem[]{} Zoccali, M., Renzini, A., Ortolani, S., et al. 2003, A\&A, 399, 931

\end{thebibliography}
\end{document}